\documentclass[compsoc,conference,a4paper,10pt,times]{IEEEtran}
\IEEEoverridecommandlockouts
\usepackage{cite}
\usepackage{amsmath,amssymb,amsfonts}
\usepackage{algorithmic}
\usepackage{graphicx}
\usepackage{textcomp}
\usepackage{bmpsize}
\usepackage{xcolor}
\usepackage{lipsum}
\usepackage{booktabs}
\usepackage{multirow}
\usepackage{makecell}
\usepackage{balance}
\usepackage[colorlinks=true,urlcolor=black]{hyperref}
\def\BibTeX{{\rm B\kern-.05em{\sc i\kern-.025em b}\kern-.08em
    T\kern-.1667em\lower.7ex\hbox{E}\kern-.125emX}}
\begin{document}

\title{Evidence-Based Threat Modeling for ICS}

 \author{\IEEEauthorblockN{Can \"{O}zkan}
 \IEEEauthorblockA{\textit{KU Leuven - COSIC} \\
 Kasteelpark Arenberg 10\\
 3001 Heverlee, Belgium \\
 can.ozkan@kuleuven.be}
 \and
 \IEEEauthorblockN{Dave Singel\'{e}e}
 \IEEEauthorblockA{\textit{KU Leuven - COSIC} \\
 Kasteelpark Arenberg 10\\
 3001 Heverlee, Belgium \\
 dave.singelee@kuleuven.be}
}

\maketitle

\begin{abstract}
ICS environments are vital to the operation of critical infrastructure such as power grids, water treatment facilities, and manufacturing plants. However, these systems are vulnerable to cyber attacks due to their reliance on interconnected devices and networks, which could lead to catastrophic failures. Therefore, securing these systems from cyber threats becomes paramount. In this context, threat modeling plays an essential role. Despite the advances in threat modeling, the fundamental gap in the state-of-the-art is the lack of a systematic methodology for identifying threats in ICS comprehensively. Most threat models in the literature (i) rely on expert knowledge, (ii) only include generic threats such as spoofing, tampering, etc., and (iii) these threats are not comprehensive enough for the systems in question. To overcome these limitations, we propose a novel evidence-based methodology to systematically identify threats based on existing CVE entries of components and their associated fundamental weaknesses in the form of CWE entries -- namely, CVE-CWE pairs -- and thereby generate a comprehensive threat list. Furthermore, we have implemented our methodology as a ready-to-use tool and have applied it to a typical SCADA system to demonstrate that our methodology is practical and applicable in real-world settings.

\end{abstract}

\begin{IEEEkeywords}
Threat modeling, security, ICS, ICS security, SCADA
\end{IEEEkeywords}

\section{Introduction}\label{sec:introduction}

Industrial control system (ICS) is a general term that encompasses several types of control systems, including SCADA systems, distributed control systems (DCS), and other control system configurations such as Programmable Logic Controllers (PLC) often found in the industrial sectors and critical infrastructures\cite{stouffer2011guide}. These systems are paramount in automating industrial processes, enabling real-time monitoring, data acquisition, and process control, and optimizing efficiency, safety, and reliability. It collects sensor data, processes it in real time, and sends commands to machines or equipment to control their operations, such as temperature, machinery movement, or chemical processing.

A Supervisory Control and Data Acquisition (SCADA) system\cite{daneels1999scada} is an example of an ICS system that monitors and supervises industrial processes among various sectors, including manufacturing facilities, oil production and processing, pharmaceuticals, energy, water treatment, and distribution \cite{sami2019scada}. SCADA systems combine data acquisition and transmission systems with HMI software to provide a centralized monitoring and control system for multiple process inputs and outputs\cite{stouffer2011guide}. These systems gather field information, send it to a central computer facility, and present it to the operator in graphical or textual form through the HMI component, enabling the operator to supervise and monitor an entire system from a central location in real-time.

ICS networks were typically isolated in the past, and cyber security was not a primary concern during the early stages of system development. However, due to the increasing complexity and interconnected structure of IT and ICS systems, attack surfaces have grown significantly for adversaries. Although this convergence of IT/ICS has contributed to efficiency, ease, and innovation, it has also increased the attack surface, thereby making these systems prime targets of cyber threats\cite{murray2017convergence}.

This increasing number of emerging cyber threats poses the need to secure ICS, particularly SCADA systems. SCADA systems manage power grids, water treatment facilities, and transportation networks—in other words, societies' critical infrastructures. Successful attacks can severely damage these infrastructures, resulting in anything from prolonged service outages to public safety and even national security.

In this context, threat modeling is paramount and plays an essential role in protecting ICSs. Threat modeling is the practice of systematically identifying, analyzing, and mitigating potential threats within a system, application, or process. By assessing security threats during the design and development phases, organizations can improve their security posture, reduce the likelihood of successful attacks, and ensure compliance with security requirements \cite{shostack2014threat}. Notable methodologies such as STRIDE\cite{shostack2014threat}, DREAD, PASTA\cite{ucedavelez2015risk}, OCTAVE\cite{alberts2003introduction}, and attack trees\cite{schneier1999attack} provide structured approaches for threat modeling, each offering unique perspectives for analyzing threats \cite{myagmar2005threat}. Hence, organizations can design more secure systems and implement proper countermeasures to protect against such threats.


Threat identification is the fundamental step and central concept in all threat modeling methodologies such as STRIDE, PASTA, OCTAVE, and attack trees. Each methodology considers threat identification an essential step in listing threats for a system in question.

However, a fundamental limitation of existing threat modeling methodologies is that although they are needed to identify threats, they provide limited guidance on how to do so systematically. Most threat models in the literature (i) rely on expert knowledge, (ii) mostly only include generic threats such as spoofing, tampering, etc., and (iii) the threats resulting from these methods are not comprehensive enough for the systems
in question. This lack of specificity can lead to variability in threat results, as different individuals or teams might generate different threat lists based on their experiences and perspectives. Therefore, the threats may not be complete and critical threats might be overlooked, or less significant ones might be emphasized too much.



Our contribution is twofold. First, we have developed a methodology that provides a structured and systematic approach to threat identification. This methodology offers clear steps - first, analyzing historical vulnerability information for each component; second, obtaining underlying weaknesses for each vulnerability (CVE, CWE pairs); third, de-duplicating CWE entries, thereby deducing threats - to address the limitations associated with existing threat modeling methodologies. Second, we have developed a ready-to-use tool that implements the proposed methodology, allowing practitioners to apply it in real-world settings easily. The tool automates the key aspects of our methodology, enhancing threat coverage and accuracy in threat elicitation. Together, the methodology and the accompanying tool provide a practical solution for threat modelers, engineers, and operators in the ICS domain.

The rest of the paper is organized as follows: In Section \ref{sec:literature_review}, we present a comprehensive review of the related work, highlighting gaps our approach aims to address. Section \ref{sec:main_concept} discusses the main reasoning behind the proposed methodology. In Section \ref{sec:evidence_based_methodology}, we provide a detailed explanation of the methodology itself. In Section \ref{sec:software_tool}, we present the tool we have developed. Section \ref{sec:discussion} discusses the integration of our methodology with existing ones and the limitations of our methodology. Section \ref{sec:case_study} presents a case study in which we apply our methodology to a sample SCADA system, demonstrating its practical implications in real-world settings. Finally, Section \ref{sec:conclusion} concludes the paper with a discussion.
\section{Related work}\label{sec:literature_review}

\subsection{Existing Threat Modeling Methodologies}

Researchers have proposed various approaches to address unique threat modeling challenges in ICS environments. These approaches generally build on existing methodologies like STRIDE, OCTAVE, and attack tree, while others have proposed new frameworks or combined different methods to address ICS threat modeling needs.

Several papers have applied the STRIDE methodology to ICS environments, ranging from smart grids, synchrophasor-based synchronous islanding systems, smart meters, Advanced Metering Infrastructure, energy-related applications, and agricultural systems.\cite{ahn2021security, li2021vision, khan2017stride, girdhar2021hidden, rouland2021specification, haider2019threat, kim2022stride, ferrer2017principles, chen2019determining, al2021stride }. The fundamental issue in literature implementing STRIDE is that threats are too generic, limited to such as spoofing, tampering, and buffer overflows, and are based on expert knowledge.

Additionally, Zografopoulos et al. \cite{zografopoulos2021security} applied the OCTAVE methodology to integrated transmission and distribution power systems and considered three threats. Furthermore, we have seen that attack trees were implemented. Martins et al. \cite{martins2015towards} propose using attack trees with generic ten threats to the Wireless Railway Temperature Monitoring System. Stellios et al. \cite{stellios2021assessing} proposed a risk-based methodology combining cyber and physical interactions. It utilizes attack trees to represent potential attack paths. The risk of each attack path is then assessed based on its likelihood and impact. The fundamental issue in this category is that threat models performed with the attack tree mostly focused on risk assessment.


Several papers have performed an ad-hoc threat assessment. Zografopoulos et al. \cite{zografopoulos2021cyber} applied threat modeling to cyber-physical energy systems, identifying nine general threats, such as man-in-the-middle (MiTM), DoS, and engineering workstation compromise. Radoglou-Grammatikis et al.\cite{radoglou2019attacking} focused on the threats in the IEC-104 protocol used in SCADA systems, identifying threats such as unauthorized access and DoS attacks. They identified 13 threats to the IEC-104 protocol. By Liu et al. \cite{liu2015collaborative}, a threat model with nine cyber threats was performed to identify attack points, specifically targeting data manipulation and command tampering for AMI.

Several papers have proposed new methodologies or combinations of approaches in addition to existing ones. Valenza et al. \cite{valenza2022hybrid} introduced TAMELESS, an automated tool for analyzing hybrid threats that span human, physical, and cyber domains. However, the threats identified remained limited to generic categories such as unauthenticated access and physical damage, i.e., physical threats to carry out a cyber-attack (e.g., installing malware on a network switch). Stellios et al. \cite{stellios2021assessing} proposed a risk-based methodology combining cyber and physical attack paths, utilizing attack trees to represent potential attack paths. Their goal is to assess the risk of attack paths of interacting nodes towards a critical target. The proposed methodology by Schlegel et al. \cite{schlegel2015structured} requires users to populate the threat model with a list of threats. The fundamental problem -no systematic, comprehensive list of threats- of threat modeling is still inherent. Framework by Zahid et al. \cite{zahid2023threat} performed a threat modeling depending on the MITRE ATT\&CK matrix to identify possible threats to smart firefighting. Then, they map relevant threats to smart firefighting systems based on expert judgment.

Despite the advances in threat modeling and its applications in the ICS, the fundamental gap in the literature is the lack of a systematic method for identifying threats comprehensively. The existing literature is relatively generic for threat catalog, relied on expert knowledge, or focused on risk assessment rather than comprehensive threat identification in the ICS domain. This raises concerns about the comprehensiveness, granularity, and accuracy of generated threats. Relying on generic threats and expert input can result in an incomplete threat model. Therefore, the literature review paper pointed out a clear need for a systematic methodology that overcomes such limitations.

\subsection{Threat Modeling Tools}

Most threat modeling tools focus on software development and are not intended for accurately modeling threats to industrial networks. 

Microsoft designed the Microsoft Threat Modeling Tool\cite{Microsoft_TM_Tool} to support the STRIDE methodology. This user-friendly tool integrates well with Microsoft's development environment and technologies and heavily supports Microsoft products. However, Fla et al. \cite{flaa2021tool} developed a custom Smart Grid template for the TMT, tailored to the specific components and processes in smart grid environments. This template helps asset owners systematically model threats and classify them according to risk levels.

OWASP Threat Dragon\cite{OWASP_Threat_Dragon} is an open-source tool that provides a simple interface for creating threat models. However, it does not offer any threat list by default and relies solely on the users' expert knowledge. Additionally, ThreatModeler\cite{Threat_Modeler} and IriusRisk\cite{IriusRisk} are commercial threat modeling tools. IriusRisk allows users to perform only one threat modeling for a demo purpose. However, the demo version does not provide threat lists for ICS environments. Therefore, it is essential to have a tool that systematically identifies ICS threats.

\section{Main Concept}\label{sec:main_concept}

In this section, we discuss the central concept behind our evidence-based methodology. We explain the origin of the CVE-CWE pairs concept and how these concepts are interconnected and can help us systematically identify threats in a given ICS environment. 

The Common Vulnerabilities and Exposures (CVE) \cite{CVE} system is a public database of standardized identifiers for known security vulnerabilities and provides a unique identifier (a CVE ID) for each discovered vulnerability. This enables organizations, researchers, and security professionals to track, share, and address vulnerabilities in a standardized way. The description of each CVE entry contains a brief overview of the vulnerability, its possible consequences, and links to associated advisories, fixes, and potential remediation. This approach standardizes efficient information sharing among security professionals.

A typical CVE entry contains a description summarizing the vulnerability, explaining how it affects software or hardware systems and the potential impact. It also specifies the affected products and versions, allowing organizations to determine if they are at risk. CVE entries often include severity ratings as a Common Vulnerability Scoring System (CVSS) score; although CVSS itself is proposed as a threat modeling methodology, it is contradicted by Adam Shostack and proposed to be used only as a scoring system \cite{adam_blog}. They may also provide exploit details step by step, including attack vectors and the availability of exploits in the wild. In addition to a single CVE entry, more can be learned by analyzing bulk CVE entries, and it will be discussed later in this section. 


CWE \cite{CWE}, on the other hand, is a catalog of standard software and hardware weakness types that can lead to security vulnerabilities. Each entry in the CWE list is assigned a unique identifier (e.g., CWE-119) and includes a description, potential consequences, examples, and mitigation guidance. 
Despite serving different purposes, CVE and CWE share a relationship in addressing security issues. CWE links vulnerabilities to their fundamental security weaknesses. In other words, while CVE entries list specific vulnerabilities in products or components, CWE entries describe the fundamental weaknesses that lead to such vulnerabilities. Therefore, each CVE entry can be associated with one or more CWE identifiers defining the root cause of a vulnerability. For instance, a CVE entry describing a buffer overflow vulnerability in a PLC might be linked to CWE-119, improper restriction of operations within the bounds of a memory buffer. In this case, improper restrictions within a memory buffer is the fundamental weakness and it leads to buffer overflow vulnerabilities in PLC and SCADA systems. Another example is a CVE entry representing a cross-site scripting (XSS) vulnerability in a PLC web user interface. This vulnerability is linked to CWE-79, improper neutralization of input during web page generation. In this case, the fundamental weakness that leads to XSS is that a product does not neutralize or improperly neutralize the user input used as an output during the web page generation.

In addition, the CWE database contains 1,363 entries at the time of writing. It means that it provides a highly granular view of software and hardware weaknesses. This extensive catalog allows security professionals and practitioners to rigorously classify vulnerabilities within systems and applications to their fundamental weaknesses.

From the threat modeling point of view, threats principally exploit vulnerabilities to carry out attacks or harm systems. More specifically, vulnerabilities are flaws or weaknesses that threats can exploit, and weaknesses can lead to vulnerabilities, which, when targeted by adversaries, threats can be realized through the exploitation of those vulnerabilities. For instance, attackers exploit buffer overflow vulnerabilities to inject and execute arbitrary code on the target system. In this case, the threat is remote code execution on the target by threat actors. The vulnerability is the buffer overflow. The underlying weakness that leads to this vulnerability, thereby realizing the threat, is improper restrictions within the bounds of a memory buffer, CWE-119. Therefore, if we know the weakness that allows this vulnerability, the threat can be prevented. Thus, it is essential to obtain weaknesses in this context. Put simply, by identifying weaknesses, we can deduce and cover threats. 

Knowing the relationship between vulnerability, weakness, and threat, we can now focus on the principal concept of our methodology. As vulnerabilities exploit underlying weaknesses and flaws in systems and components, examining prior CVE entries from similar devices of the same component type (e.g., PLC) can help anticipate upcoming vulnerabilities and uncover recurring issues and common attack patterns that threat actors exploit. This understanding facilitates predicting where future vulnerabilities might occur. However, we think that analyzing solely vulnerability information is not enough, and combining this with the weakness information is more valuable in that if we address the underlying weakness, the upcoming vulnerabilities that can emerge due to this weakness can be prevented in addition to deducing threat information as discussed in the previous paragraph. By systematically studying this approach in the form of CVE-CWE pairs for a component, we can learn all the weaknesses to date from the prior vulnerabilities that have affected the component. This is where the evidence-based concept comes from. For this reason, if CVE-CWE pairs are analyzed and documented, then threats to a component can be comprehensively derived without the need for expert knowledge. It can then be used as a methodology that will be discussed in Section \ref{sec:evidence_based_methodology} or a sub-methodology in existing methodologies in order to identify threats for components systematically.

Consequently, our proposed evidence-based methodology's main elements are primarily fundamental weaknesses learned from bulk historical vulnerability data. Analyzing this historical vulnerability data and then obtaining weaknesses can yield valuable information in deducing threats, and this CWE list results in a granular threat list for each component. It ultimately solves the gap in the literature as it provides more granularity than generic threats such as tampering, spoofing, and SQL injections and eliminates the need for expert knowledge.

\section{Our Novel Evidence-based Threat Modeling Methodology}\label{sec:evidence_based_methodology}
We now introduce our new threat modeling methodology and dissect the steps to provide a comprehensive understanding of its application. Our methodology builds upon existing frameworks by introducing novel elements that enhance threat identification in complex ICS systems.

\textbf{Define Scope and Assets:}
This phase establishes the scope and determines what components will be analyzed in a system. It identifies the key components that need to be protected. These components include hardware, software, or any component that has previous CVE entries. A sample output in this step is that the threat modeling scope includes two PLCs, one SCADA, one actuator, two sensors, and one Windows server to deploy the SCADA server. Furthermore, libraries and other components can be added to your scope, depending on how deep a threat model it would be. For instance, the JUnit library can be added to the scope and asset list. The components in the scope will be the input of the next phase, threat identification. Contrary to STRIDE, our methodology does not require a data flow diagram of components as there are no CVE entries for data in transit, to the best of our knowledge.

\textbf{Threat Identification:}
This is the fundamental step and our major contribution to the threat modeling process for systematic and exhaustive threat elicitation. In this phase, our evidence-based methodology is leveraged to identify threats to assets and components. First, historical CVE entries for a component in the scope are analyzed. Second, fundamental weakness(es) that led to this vulnerability, namely, its CWE entry, are listed for each vulnerability. Third, duplicated CWE entries are eliminated. Therefore, the threat modeler has a list of unique weaknesses that are applicable to the component. Then, this systematic approach is applied to each component in the scope. As a result, the threat modeler has a list of weaknesses that an adversary can exploit and realize the threat. 


\textbf{Analyze and Prioritize Threats:} The objective of this phase is to prioritize the threats identified in the previous phase. Having analyzed CVE entries for a component, prioritizing the threats that occur most repeatedly can be a strategic approach as these weaknesses led to the most number of vulnerabilities in the component. Prioritizing these high-occurrence weaknesses allows security teams to focus their resources on addressing the most widespread threats first. Our software tool automatically displays the top 5 threats for each component. Therefore, organizations and threat modelers can prioritize those threats. 

\textbf{Mitigation of Threats:} The objective of this phase is to propose and plan countermeasures to reduce or eliminate identified risks. Each CWE value often comes with potential mitigation techniques to address the weaknesses. By examining these suggested mitigation techniques, organizations can implement solutions to remediate threats in their systems. This information enables security teams to apply patches, update configurations, secure software development, or adopt best practices.

\textbf{Validate and Verify Mitigation:} When threat modeling is conducted during the system design phase, the identified threats serve as essential inputs for defining the security requirements. This proactive approach ensures that potential vulnerabilities are addressed early, allowing security measures to be integrated seamlessly into the system architecture. Conversely, if threat modeling is applied to a system that is already in production, the process shifts from defining requirements to verifying that mitigations have been effectively implemented. In this context, other mechanisms such as penetration testing, vulnerability scanning, code review, fuzz testing, and design review are employed to ensure that the implemented mitigation addresses and fixes the identified threats.

\section{Software Tool\label{sec:software_tool}}
We have developed a tool to automate our threat modeling methodology. It provides a framework for visualizing system components and automatically generates threats based on our evidence-based methodology discussed in Sections 3 and 4. Furthermore, our tool offers user-friendly graphical user interfaces that allow users to input system components such as PLC, SCADA, actuator, sensor, etc., and help identify threats without needing deep security expertise. Automation of the proposed methodology significantly reduces the time and effort required for threat modeling, making the process easier, consistent, and systematic. Moreover, it not only supports threat identification in ICS networks, but also IT networks, although its main purpose is to elicit threats in ICS networks. 

To implement our methodology, we utilize two main APIs (Application Programming Interface). First, we leveraged the MitreCVE API \cite{CVE_API}. This API retrieves all CVEs for a package/keyword from the CVE MITRE database. In our implementation, the main use of this API is to retrieve all the CVE entries that affect a component. However, the main disadvantage of this API is that it does not provide a CWE entry for each CVE value. For this reason, we need another API to achieve this and provide feedback to MITRE corporation about this issue. In order to obtain the CWE entry for each CVE entry, we utilized the NIST Vulnerability API \cite{NIST_API}. The NIST Vulnerability API is used to easily retrieve information on a single CVE or a collection of CVEs from the NVD. However, the limitation of this API is that NIST firewall rules put in place to prevent denial of service attacks can thwart your application if it exceeds a predetermined rate limit. The public rate limit (without an API key) is 5 requests in a rolling 30-second window; the rate limit with an API key is 50 requests in a rolling 30-second window \cite{NIST_Threshold}. Therefore, requesting and using an API key significantly raises the number of requests that can be made in a given time frame, and it is recommended to obtain an API key in order to speed up the implementation.

Now that we have discussed the APIs utilized in our implementation, let us zoom in on the details of the tool. Users input their assets and system components, including hardware and software components that require protection on the initial screen, as shown in Figure 1. Developed using Python, the tool provides a drop-down menu listing common ICS components such as PLC, SCADA, HMI, sensor, actuator, RTU, and IED. If the component is not listed in the pre-built drop-down menu, the user can manually input it under the custom component name along with its description. This information is then visualized using rectangle entities as it will be demonstrated in the next section. By clicking the “Analyze” button, the tool applies the proposed methodology to automatically identify threats and generate a list of threats for each component. The tool then shows the top 5 occurrence threats using the matplotlib library in a pie chart, and the generated charts can be used for prioritization purposes. Furthermore, by clicking the “Show Results” button, the user can list either all threats or the top five threats for each component in the scope, which will be demonstrated in the next section in detail, where we apply our methodology and tool to a sample SCADA system. 

The tool normally needs to connect to the CVE MITRE API and NIST Vulnerability API for each threat model, which takes considerable time due to NIST rate limiting and the possibility of bulk CVE entry for each component. For instance, there exist more than 1200 CVE entries for only SCADA software. In order to overcome this limitation and ultimately speed up the threat generation process, we compiled all the CVE and CWE pairs for common ICS components in a Python dictionary in the source code. The tool offers a feature to update the CVE and CWE pairs by using the “Update Threat List” button, as can be seen in Figure 1. The threat library should be periodically updated to update the threat list. 

\begin{figure}
  \includegraphics[width=\linewidth]{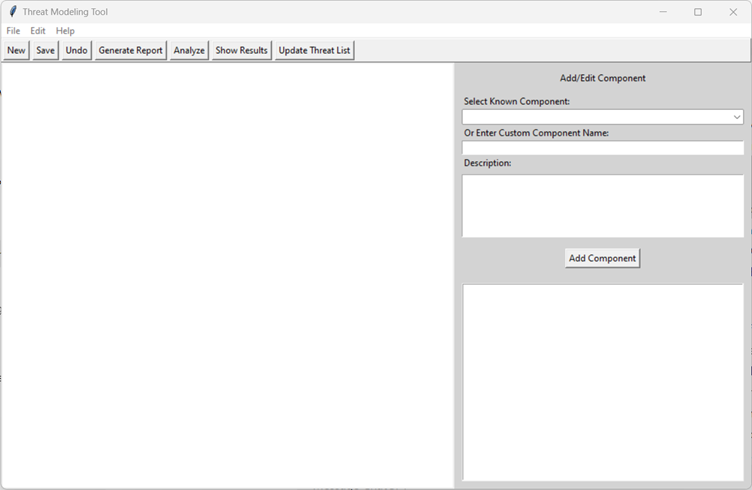}
  \caption{Initial Screen of the Tool}
  \label{fig:boat1}
\end{figure}
\section{Discussion\label{sec:discussion}}
\subsection{Integration into Existing Threat Modeling Methodologies}
Threat identification is the fundamental step in threat modeling methodologies such as PASTA, OCTAVE, STRIDE, and attack trees. However, existing methodologies lack concrete and systematic methods for identifying threats, and they are mostly based on brainstorming and expert knowledge, as discussed in Sections \ref{sec:introduction} and \ref{sec:literature_review}. Therefore, the threat identification phase of our evidence-based methodology can be employed as a sub-methodology for the threat identification phase in existing methodologies. 

\subsection{Limitations of Our Methodology}
While our proposed methodology offers significant advantages in enhancing threat identification, it is essential to also acknowledge its limitations. In this subsection, we discuss the potential constraints and areas where the methodology may require further improvement.

The major drawback is that although CWE offers a grouping mechanism, the NIST Vulnerability API to fetch CWEs does not provide any grouping mechanism. For instance, MITRE groups CWE-120, CWE-124, CWE-125, and some other related CWEs to memory buffer errors; however, the NIST API does not offer this grouping mechanism. Feedback has been sent to NIST. Furthermore, a minor criticism regarding MITRE CWE grouping is that although CWE-119, Improper Restriction of Operations within the Bounds of a Memory Buffer, is a memory buffer error, it is not available in the memory buffer errors group. Our case study, which will be discussed in Section 5, has revealed that CWE-119, a memory error issue, is the most occurring CWE entry in PLCs, and it does not belong to the memory error issue group. For this reason, there is no grouping in our tool, such as data validation issues or memory buffer errors.

Second, CWE focuses solely on technical weaknesses; namely, it primarily addresses technical weaknesses in software and hardware. It does not cover non-technical vulnerabilities such as process failures, organizational issues, or social engineering threats, which can result in security issues. In addition, as CVE and CWE entries are mostly related to cyber-security, our methodology is solely focused on the cyber-security aspect and does not directly consider privacy and safety issues in ICS. As some cyber-security incidents can result in the loss of human life, we indirectly consider and address safety issues.
\section{Case Study on a Typical SCADA Network}\label{sec:case_study}

In this section, we apply our evidence-based threat methodology and tool to a typical SCADA network to determine its applicability in real-world settings. The case study aims to evaluate our methodology and tool in identifying and addressing threats to SCADA systems by systematically examining their vulnerabilities and inherent fundamental weaknesses.

SCADA networks typically consist of several key components that work together. Programmable Logic Controllers (PLCs) and Remote Terminal Units (RTUs) are controllers and play a crucial role in connecting with machinery and sensors for controlling purposes. PLCs are devices based on microprocessors that carry out control tasks according to input signals. Similarly, RTUs gather data from sensors and send it to the SCADA master unit, the control central system on which the SCADA software is deployed. Additionally, sensors and actuators play a vital role, with sensors gauging physical parameters such as temperature and pressure and actuators executing tasks such as opening valves or initiating motors. The control center collects and logs the information gathered by the field devices. Information is then displayed to the HMI, and actions may be generated based on detected events. Despite the availability of other components, such as data historians, we keep the scope simple in our case study for simplicity. Therefore, the scope of threat modeling in our case study concludes a PLC, an RTU, a SCADA software, a sensor, and an actuator, as seen in Figure 2.

\begin{figure}
  \includegraphics[width=\linewidth]{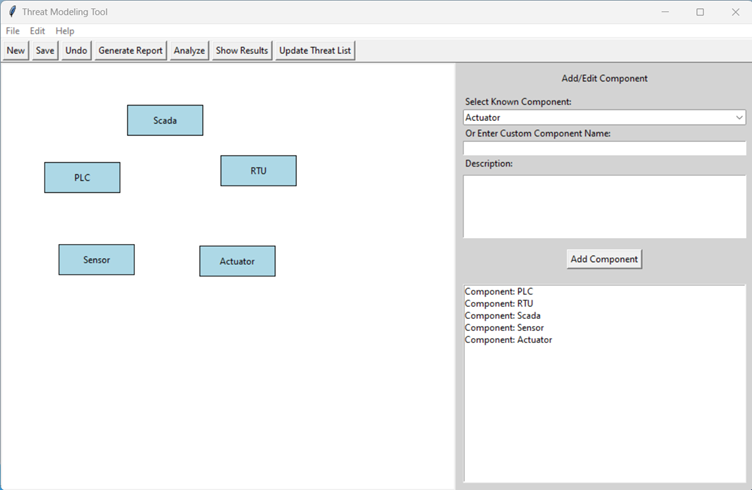}
  \caption{Scope and System Components}
  \label{scopeandsystemcomponents}
\end{figure}

Now that the scope has been identified, we can proceed with the threat identification phase. Once the "Analyze" button is pressed, the tool starts analyzing the threats for each component and shows the top 5 most frequently occurring CWE values for each component to the screen by default. Figure 3 shows that CWE-119, Improper Restriction of Operations within the Bounds of a Memory Buffer, is the top threat for PLC, with 9\% of 213 PLC CVE entries, followed by CWE-287, improper authentication with 8\%. The five most occurring threats to PLCs are as follows: CWE-119: Improper Restriction of Operations within the Bounds of a Memory Buffer, CWE-287: Improper Authentication, CWE-400: Uncontrolled Resource Consumption, CWE-306: Missing Authentication for Critical Function, CWE-20: Improper Input Validation. It can be concluded that those threats are the most occurring threats to PLCs. Thus, it is logical to prioritize such threats first. 

\begin{figure}
  \includegraphics[width=\linewidth]{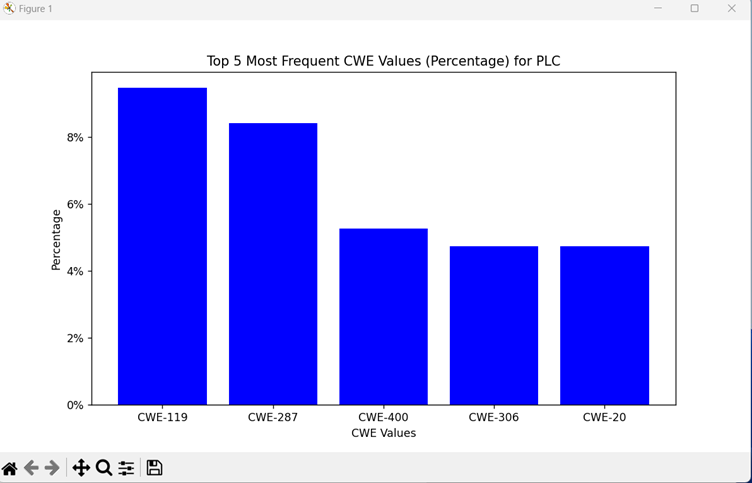}
  \caption{The Top 5 Threats for PLC}
  \label{top5threats}
\end{figure}

By clicking "Show Results," users can select all threats or the top five threats. All PLC threats are shown in Figure 4. The findings are valuable. The tool identified 60 threats, in total, for PLC and are as follows: CWE-121, CWE-125, CWE-384, CWE-294, CWE-319, CWE-312, CWE-703, CWE-676, CWE-798, CWE-306, CWE-404, CWE-494, CWE-326, CWE-416, CWE-415, CWE-284, CWE-552, CWE-347, CWE-345, CWE-434, CWE-22, CWE-425, CWE-400, CWE-522, CWE-532, CWE-787, CWE-401, CWE-672, CWE-287, CWE-427, CWE-23, CWE-755, CWE-770, CWE-20, CWE-863, CWE-94, CWE-476, CWE-119, CWE-665, CWE-120, CWE-754, CWE-307, CWE-77, CWE-862, CWE-668, CWE-201, CWE-352, CWE-290, CWE-78, CWE-353, CWE-79, CWE-200, CWE-327, CWE-662, CWE-255, CWE-254, CWE-399, CWE-16, CWE-310, CWE-295. Some notable threats for PLCs are Stack-based Buffer Overflow, Out-of-bounds Read, session fixation, Authentication Bypass by Capture-replay, Cleartext Transmission of Sensitive Information, Cleartext Storage of Sensitive Information, Improper Check or Handling of Exceptional Conditions, Use of Potentially Dangerous Function, Use of Hard-coded Credentials, Missing Authentication for Critical Function, Improper Resource Shutdown or Release, Download of Code Without Integrity Check, Inadequate Encryption Strength, Use After Free, Double Free, Improper Access Control, Files or Directories Accessible to External Parties, Improper Verification of Cryptographic Signature, Insufficient Verification of Data Authenticity, Unrestricted Upload of File with Dangerous Type, Improper Limitation of a Pathname to a Restricted Directory ('Path Traversal'). Therefore, our empirical results show that our methodology can produce concrete threats comprehensively, rather than generic spoofing and integrity threats, along with eliminating expert knowledge, and can be used in generating threats for components. Additionally, the number of threats for other components and the top 5 most occurring threats for each component are summarized in Table\ref{tab:table1} and Table\ref{tab:table2}, respectively.

\begin{figure}
  \includegraphics[width=\linewidth]{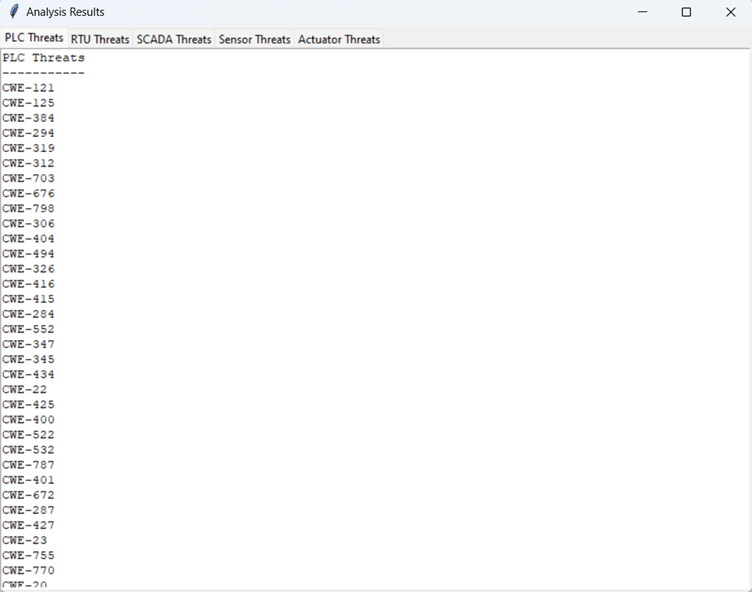}
  \caption{All PLC Threats}
  \label{allplcthreats}
\end{figure}

%

\begin{table}
        
        \caption{\# of Threats per Component}
        \label{tab:table1}
        \begin{tabular}{cccccc}
            \toprule
            \makecell[c]{Technology} & \makecell[c]{PLC} & \makecell[c]{RTU} & \makecell[c]{SCADA} & \makecell[c]{Sensor} & \makecell[c]{Actuator} \\ \midrule
            \# of Threats&  60&  29&  68&  48& 11\\
                          
            \bottomrule
        \end{tabular}
    \end{table}


\begin{table}
        
        \caption{Top 5 Threats Per Component}
        \label{tab:table2}
        \begin{tabular}{cccccc}
            \toprule
            \makecell[c]{PLC} & \makecell[c]{RTU} & \makecell[c]{SCADA} & \makecell[c]{Sensor} & \makecell[c]{Actuator} \\ \midrule
            CWE-119&  CWE-798&  CWE-119&  CWE-787& CWE-22\\ 
         CWE-287&  CWE-22&  CWE-200&  CWE-22& CWE-200\\ 
         CWE-400&  CWE-287&  CWE-20&  CWE-20& CWE-862\\  
         CWE-306&  CWE-754&  CWE-22&  CWE-264& CWE-94\\ 
         CWE-20&  CWE-200&  CWE-79&  CWE-77& CWE-732\\
                          
            \bottomrule
        \end{tabular}
    \end{table}

Overall, the results of this investigation reveal that our evidence-based methodology can produce a comprehensive set of threats for each component and eliminate expert knowledge and false positives. Also, the threats to components, particularly PLCs, such as buffer overflow and missing authentication, align with previous threat modeling results in literature. The evaluation of the tool in a real-world setting shows that it could identify threats comprehensively and provide a systematic solution for the identified gap. The case study makes this methodology practical and applicable in real-world settings. For this reason, our evidence-based methodology can be used to determine threats by itself or as a sub-methodology in other threat modeling methodologies. 

\section{Conclusion}\label{sec:conclusion}

Threat modeling is an exercise to help identify threats early in the design process, thereby reducing overall costs and improving security. However, existing threat modeling methodologies lack a systematic methodology for identifying threats in specific environments, such as, for example, ICS. These methodologies are mostly based on generic threats such as spoofing, tampering, etc., and often require expert knowledge. The goal of this paper is to propose a novel systematic methodology for identifying threats in ICS systems. We proposed an evidence-based methodology in order to systematically identify threats that are based on analyzing historical vulnerability information along with underlying weaknesses (CVE, CWE pairs), thereby deducing threats and discussed how our methodology solved the gap in the literature. Our methodology is accompanied by a software tool to automate our threat identification process. We demonstrated the applicability and accuracy of our methodology on a representative SCADA network. Our comprehensive findings for each component clearly indicate that our methodology works effectively; therefore, our methodology can be used to identify threats for components that have previous CVE entries and can be easily adopted in threat modeling exercises.  

\section*{Acknowledgements}
This work has been supported in part by CyberSecurity Research Flanders with reference number VR20192203, by the Energy Transition Fund of the FPS Economy of Belgium through the CYPRESS project, and by the VLAIO COOCK program through the IIoT-SBOM project.

\bibliographystyle{IEEEtran}
\bibliography{IEEEabrv,biblio.bib}

\begin{thebibliography}{10}
\providecommand{\url}[1]{#1}
\csname url@samestyle\endcsname
\providecommand{\newblock}{\relax}
\providecommand{\bibinfo}[2]{#2}
\providecommand{\BIBentrySTDinterwordspacing}{\spaceskip=0pt\relax}
\providecommand{\BIBentryALTinterwordstretchfactor}{4}
\providecommand{\BIBentryALTinterwordspacing}{\spaceskip=\fontdimen2\font plus
\BIBentryALTinterwordstretchfactor\fontdimen3\font minus \fontdimen4\font\relax}
\providecommand{\BIBforeignlanguage}[2]{{%
\expandafter\ifx\csname l@#1\endcsname\relax
\typeout{** WARNING: IEEEtran.bst: No hyphenation pattern has been}%
\typeout{** loaded for the language `#1'. Using the pattern for}%
\typeout{** the default language instead.}%
\else
\language=\csname l@#1\endcsname
\fi
#2}}
\providecommand{\BIBdecl}{\relax}
\BIBdecl

\bibitem{stouffer2011guide}
K.~Stouffer, J.~Falco, K.~Scarfone \emph{et~al.}, ``Guide to industrial control systems (ics) security,'' \emph{NIST special publication}, vol. 800, no.~82, pp. 16--16, 2011.

\bibitem{daneels1999scada}
A.~Daneels and W.~Salter, ``What is scada?'' 1999.

\bibitem{sami2019scada}
A.~Sami, ``Scada (supervisory control and data acquisition),'' 2019.

\bibitem{murray2017convergence}
G.~Murray, M.~N. Johnstone, and C.~Valli, ``The convergence of it and ot in critical infrastructure,'' 2017.

\bibitem{shostack2014threat}
A.~Shostack, \emph{Threat modeling: Designing for security}.\hskip 1em plus 0.5em minus 0.4em\relax John Wiley \& Sons, 2014.

\bibitem{ucedavelez2015risk}
T.~UcedaVelez and M.~M. Morana, \emph{Risk Centric Threat Modeling: process for attack simulation and threat analysis}.\hskip 1em plus 0.5em minus 0.4em\relax John Wiley \& Sons, 2015.

\bibitem{alberts2003introduction}
C.~Alberts, A.~Dorofee, J.~Stevens, and C.~Woody, ``Introduction to the octave approach,'' \emph{Pittsburgh, PA, Carnegie Mellon University}, pp. 72--74, 2003.

\bibitem{schneier1999attack}
B.~Schneier, ``Attack trees,'' \emph{Dr. Dobb’s journal}, vol.~24, no.~12, pp. 21--29, 1999.

\bibitem{myagmar2005threat}
S.~Myagmar, A.~J. Lee, and W.~Yurcik, ``Threat modeling as a basis for security requirements,'' 2005.

\bibitem{ahn2021security}
B.~Ahn, T.~Kim, S.~C. Smith, Y.-W. Youn, and M.-H. Ryu, ``Security threat modeling for power transformers in cyber-physical environments,'' in \emph{2021 IEEE Power \& Energy Society Innovative Smart Grid Technologies Conference (ISGT)}.\hskip 1em plus 0.5em minus 0.4em\relax IEEE, 2021, pp. 1--5.

\bibitem{li2021vision}
K.~Li, A.~Rashid, and A.~Roudaut, ``Vision: security-usability threat modeling for industrial control systems,'' in \emph{Proceedings of the 2021 European Symposium on Usable Security}, 2021, pp. 83--88.

\bibitem{khan2017stride}
R.~Khan, K.~McLaughlin, D.~Laverty, and S.~Sezer, ``Stride-based threat modeling for cyber-physical systems,'' in \emph{2017 IEEE PES Innovative Smart Grid Technologies Conference Europe (ISGT-Europe)}.\hskip 1em plus 0.5em minus 0.4em\relax IEEE, 2017, pp. 1--6.

\bibitem{girdhar2021hidden}
M.~Girdhar, J.~Hong, H.~Lee, and T.-J. Song, ``Hidden markov models-based anomaly correlations for the cyber-physical security of ev charging stations,'' \emph{IEEE Transactions on Smart Grid}, vol.~13, no.~5, pp. 3903--3914, 2021.

\bibitem{rouland2021specification}
Q.~Rouland, B.~Hamid, and J.~Jaskolka, ``Specification, detection, and treatment of stride threats for software components: Modeling, formal methods, and tool support,'' \emph{Journal of Systems Architecture}, vol. 117, p. 102073, 2021.

\bibitem{haider2019threat}
M.~H. Haider, S.~B. Saleem, J.~Rafaqat, and N.~Sabahat, ``Threat modeling of wireless attacks on advanced metering infrastructure,'' in \emph{2019 13th International Conference on Mathematics, Actuarial Science, Computer Science and Statistics (MACS)}.\hskip 1em plus 0.5em minus 0.4em\relax IEEE, 2019, pp. 1--6.

\bibitem{kim2022stride}
K.~H. Kim, K.~Kim, and H.~K. Kim, ``Stride-based threat modeling and dread evaluation for the distributed control system in the oil refinery,'' \emph{ETRI Journal}, vol.~44, no.~6, pp. 991--1003, 2022.

\bibitem{ferrer2017principles}
B.~R. Ferrer, S.~O. Afolaranmi, and J.~L.~M. Lastra, ``Principles and risk assessment of managing distributed ontologies hosted by embedded devices for controlling industrial systems,'' in \emph{IECON 2017-43rd Annual Conference of the IEEE Industrial Electronics Society}.\hskip 1em plus 0.5em minus 0.4em\relax IEEE, 2017, pp. 3498--3505.

\bibitem{chen2019determining}
Y.-T. Chen and C.-C. Huang, ``Determining information security threats for an iot-based energy internet by adopting software engineering and risk management approaches,'' \emph{Inventions}, vol.~4, no.~3, p.~53, 2019.

\bibitem{al2021stride}
M.~R. Al~Asif, K.~F. Hasan, M.~Z. Islam, and R.~Khondoker, ``Stride-based cyber security threat modeling for iot-enabled precision agriculture systems,'' in \emph{2021 3rd International Conference on Sustainable Technologies for Industry 4.0 (STI)}.\hskip 1em plus 0.5em minus 0.4em\relax IEEE, 2021, pp. 1--6.

\bibitem{zografopoulos2021security}
I.~Zografopoulos, C.~Konstantinou, N.~G. Tsoutsos, D.~Zhu, and R.~Broadwater, ``Security assessment and impact analysis of cyberattacks in integrated t\&d power systems,'' in \emph{Proceedings of the 9th workshop on modeling and simulation of cyber-physical energy systems}, 2021, pp. 1--7.

\bibitem{martins2015towards}
G.~Martins, S.~Bhatia, X.~Koutsoukos, K.~Stouffer, C.~Tang, and R.~Candell, ``Towards a systematic threat modeling approach for cyber-physical systems,'' in \emph{2015 Resilience Week (RWS)}.\hskip 1em plus 0.5em minus 0.4em\relax IEEE, 2015, pp. 1--6.

\bibitem{stellios2021assessing}
I.~Stellios, P.~Kotzanikolaou, and C.~Grigoriadis, ``Assessing iot enabled cyber-physical attack paths against critical systems,'' \emph{Computers \& Security}, vol. 107, p. 102316, 2021.

\bibitem{zografopoulos2021cyber}
I.~Zografopoulos, J.~Ospina, X.~Liu, and C.~Konstantinou, ``Cyber-physical energy systems security: Threat modeling, risk assessment, resources, metrics, and case studies,'' \emph{IEEE Access}, vol.~9, pp. 29\,775--29\,818, 2021.

\bibitem{radoglou2019attacking}
P.~Radoglou-Grammatikis, P.~Sarigiannidis, I.~Giannoulakis, E.~Kafetzakis, and E.~Panaousis, ``Attacking iec-60870-5-104 scada systems,'' in \emph{2019 IEEE World Congress on Services (SERVICES)}, vol. 2642.\hskip 1em plus 0.5em minus 0.4em\relax IEEE, 2019, pp. 41--46.

\bibitem{liu2015collaborative}
X.~Liu, P.~Zhu, Y.~Zhang, and K.~Chen, ``A collaborative intrusion detection mechanism against false data injection attack in advanced metering infrastructure,'' \emph{IEEE Transactions on Smart Grid}, vol.~6, no.~5, pp. 2435--2443, 2015.

\bibitem{valenza2022hybrid}
F.~Valenza, E.~Karafili, R.~V. Steiner, and E.~C. Lupu, ``A hybrid threat model for smart systems,'' \emph{IEEE Transactions on Dependable and Secure Computing}, vol.~20, no.~5, pp. 4403--4417, 2022.

\bibitem{schlegel2015structured}
R.~Schlegel, S.~Obermeier, and J.~Schneider, ``Structured system threat modeling and mitigation analysis for industrial automation systems,'' in \emph{2015 IEEE 13th International Conference on Industrial Informatics (INDIN)}.\hskip 1em plus 0.5em minus 0.4em\relax IEEE, 2015, pp. 197--203.

\bibitem{zahid2023threat}
S.~Zahid, M.~S. Mazhar, S.~G. Abbas, Z.~Hanif, S.~Hina, and G.~A. Shah, ``Threat modeling in smart firefighting systems: Aligning mitre att\&ck matrix and nist security controls,'' \emph{Internet of Things}, vol.~22, p. 100766, 2023.

\bibitem{Microsoft_TM_Tool}
Microsoft, ``{Microsoft Threat Modeling Tool},'' \url{https://learn.microsoft.com/en-us/azure/security/develop/threat-modeling-tool}, [Online; accessed 11-October-2024].

\bibitem{flaa2021tool}
L.~H. Fl{\aa}, R.~Borgaonkar, I.~A. T{\o}ndel, and M.~G. Jaatun, ``Tool-assisted threat modeling for smart grid cyber security,'' in \emph{2021 International Conference on Cyber Situational Awareness, Data Analytics and Assessment (CyberSA)}.\hskip 1em plus 0.5em minus 0.4em\relax IEEE, 2021, pp. 1--8.

\bibitem{OWASP_Threat_Dragon}
OWASP, ``{OWASP Threat Dragon Tool},'' \url{https://owasp.org/www-project-threat-dragon/}, [Online; accessed 11-October-2024].

\bibitem{Threat_Modeler}
T.~Modeler, ``{Threat Modeler},'' \url{https://threatmodeler.com/}, [Online; accessed 11-October-2024].

\bibitem{IriusRisk}
IriusRisk, ``{IriusRisk},'' \url{https://www.iriusrisk.com/}, [Online; accessed 11-October-2024].

\bibitem{CVE}
MITRE, ``{CVE},'' \url{https://cve.mitre.org/index.html}, [Online; accessed 11-October-2024].

\bibitem{adam_blog}
A.~Shostack, ``{Threat Modeling},'' \url{https://shostack.org/resources/threat-modeling}, [Online; accessed 15-October-2024].

\bibitem{CWE}
MITRE, ``{CVE},'' \url{https://cwe.mitre.org/}, [Online; accessed 11-October-2024].

\bibitem{CVE_API}
------, ``{CVE API},'' \url{https://mitrecve.readthedocs.io/en/latest/api.html}, [Online; accessed 11-October-2024].

\bibitem{NIST_API}
NIST, ``{NIST Vulnerability API},'' \url{https://nvd.nist.gov/developers/vulnerabilities}, [Online; accessed 11-October-2024].

\bibitem{NIST_Threshold}
------, ``{NIST Threshold},'' \url{https://nvd.nist.gov/developers/start-here#:~:text=Rate%20Limits,a%20rolling%2030%20second%20window}, [Online; accessed 11-October-2024].

\end{thebibliography}
\balance
\end{document}